\documentclass{oapaper}        % options: a4, final, nocolorlinks

\usepackage{orcidlink}
\usepackage[normalem]{ulem}
\usepackage{xcolor}
\newif\ifshowedits
\showeditsfalse
\newcommand{\rev}[2]{%
  \ifshowedits
    {\color{red}\sout{#1}}{\color{blue}#2}%
  \else
    #2%
  \fi
}

\begin{document}

%\papertitle{RESEARCH ARTICLE}  % optional label above the title; delete if unwanted

\title{When quantum field theory kicks in: the relation between matter, gravity and energy}
%\righttitle{Another critical look at Inflationary Cosmology}  % running head, odd pages
%\lefttitle{Antonio Ferreiro and Alex Fleuren}                % running head, even pages

\author{Antonio Ferreiro$^{1}$}
\author{Andrea Cavallaro\orcidlink{0009-0002-1314-2342}}

\affil{$^{1}$Descartes Centre for the History and Philosophy of the Sciences and the Humanities, Utrecht University}
%\affil{$^\ast$Corresponding author: Antonio Ferreiro; \email{a.e.ferreirodeaguiar@uu.nl}}

%\license{\copyright\ The Author(s) 2026. \ccby}  % optional; delete for no licence line

\begin{abstract}
The existence of energy is the main criterion for (i) providing a common
core to the materiality of all types of classical matter and (ii) denying that the
gravitational field of General Relativity is a type of matter. Extending this
criterion to quantum field theory, though formally straightforward, carries new
conceptual consequences for the interpretation of gravity. Renormalization of the
quantum matter stress-energy tensor forces a local, purely geometric term, built
from the metric alone, into the conserved stress-energy, and this has two faces.
It grants the metric a \textit{material} significance alongside its spatio-temporal
one---the \textit{egalitarian} view; and it binds gravitational to non-gravitational
energy so tightly that the two cannot be split apart without a choice of vacuum that
a generic spacetime does not supply---the \textit{unificationist} view. The latter poses a dilemma: either gravitational energy is meaningless, but then so is the
energy of the non-gravitational quantum fields, and only classical fields remain matter; or
both are meaningful, and the metric thereby acquires the material significance of a
matter field. On neither horn does the classical divide between matter and spacetime
survive quantization.
\end{abstract}

\maketitle

\section{Introduction}
The description of phenomena in terms of \textit{fields} counts among the most far-reaching achievements of modern physics. In classical physics, electromagnetism and General Relativity are the paradigmatic examples, but fluid dynamics and heat conduction also testify to its explanatory reach. Fields are likewise the building blocks of quantum theories, such as Quantum Electrodynamics or the Landau–Ginzburg theory. The Standard Model, in particular, is often cast as a periodic table of elementary \textit{particles}, but its starting point, at least mathematically, is a Lagrangian composed of a set of classical fields\footnote{Strictly speaking, one starts from a classical, pre-quantization field theory with a Lagrangian and Euler--Lagrange equations, where bosonic fields are ordinary c-number fields and fermionic fields are Grassmann-valued.} that are then \textit{quantized}. Its mathematical formulation is provided by the theory of relativistic\footnote{By relativistic we mean Poincaré-invariant theories.} quantum fields, or Quantum Field Theory (QFT), which includes constituents of every type: electrons, photons, the Higgs boson, and so on. Whereas in classical electromagnetism the constituents would be electromagnetic fields, with the charges described by point particles, quantum electrodynamics places photons and electrons (and positrons) on the same footing, namely as quantum fields. In this sense, it renders all types of constituents and interactions of \textit{matter} in the form of fields.

It is precisely this leveling of entities that invites a further question: if matter is uniformly rendered as fields, can one naively force a correspondence between matter and fields, that is, say that matter is \textit{just} fields given by a particular QFT? The theory resists any straightforward answer, for two reasons. First, a quantum field, say $\hat{\phi}$, is not a classical quantity assigning a value at each point, but an operator-valued distribution defining an observable once smeared against a test function, so that a localized claim of the form ``the field vanishes here'' is not well-posed. It thus offers no direct grip on the matter content. Second, the algebra of fields alone does not determine the physical situation; one must also specify a state, which assigns expectation values and correlation functions to the observables, and the same field algebra can support physically very different states.\footnote{This is vividly illustrated by the fate of the particle concept. The standard route to particles builds a Fock space over a one-particle Hilbert space, yielding a distinguished vacuum and number operator; but singling out the one-particle subspace requires a frequency splitting of the classical solutions, which presupposes a timelike Killing field of the background spacetime \citep{Ashtekar:1975zn}\rev{}{.} Absent such an isometry, the admissible quantizations furnish unitarily inequivalent Fock representations, none preferred, and the particle notion fractures into mutually incommensurable variants—the Unruh effect being the familiar symptom. See \cite{Ruetsche2013-RUEIQT-3}, chs.~9--10.} Taken together, these two facts show that what counts as matter in a given QFT cannot simply be read off from its Lagrangian; the characterization of matter becomes an interpretive problem rather than a matter of inspection. In view of this, rather than addressing the broader ontological matter of what matter is according to QFT, we can start by asking: \textit{what makes a quantum field a matter field?}

In a recent work, \cite{Ferreiro:2025fii} proposed a criterion for something to be matter within a relativistic\footnote{Here relativistic is extended to general Lorentzian-metric manifolds.} classical field theory based on three ingredients: the existence of dynamics, the existence of a stress-energy tensor $T^{\mu\nu}$ and an equivalence between interaction at the level of the dynamics and of (stress-)energy exchange. In this work we aim to extend this criterion to quantum field theories. Two obstacles must be overcome. First, as noted above, quantum fields are operator-valued distributions, making the construction of composite observables such as the stress-energy tensor non-trivial. Second, the existence of inequivalent choices of quantum state requires the criterion to characterize the field independently of any particular state. The criterion proposed in Criterion~\ref{crit:2} is designed to address both problems.

An important conclusion drawn in \cite{Ferreiro:2025fii} was that the criterion at hand excluded the metric $g_{\mu\nu}$ of General Relativity from the set of matter fields of the theory (see §\ref{sec:classical}). Here we argue that the situation becomes less clear in quantum field theory once semiclassical gravity is brought into the discussion. In particular, the classical arguments against constructing a stress-energy tensor for gravity do not translate straightforwardly into quantum field theory on a classical curved spacetime. Renormalization of the quantum stress-energy tensor for non-gravitational fields introduces a local, purely geometric contribution, built from curvature terms ---dependent only on the metric--- which contributes to the total conserved stress-energy and gravitates. In this precise sense, quantum field theory \textit{does} afford a stress-energy tensor for gravity where the classical theory did not.

The presence of curvature terms in the stress-energy tensor carries a first philosophical outcome, which we call the \textit{egalitarian} reading, following \cite{LEHMKUHL200883}: the metric acquires a \textit{material} significance it lacks classically, and can no longer be assigned cleanly to the ``spacetime'' side of the spacetime/matter dichotomy. We do not, however, conclude that gravitational energy exists as an autonomous quantity, measurable in its own right. Here a second philosophical outcome, the \textit{unificationist} reading, pulls in the opposite direction. The decomposition of the total stress-energy into a ``non-gravitational'' contribution and a ``gravitational'' one is arbitrary and only possible in spacetimes endowed with suitable symmetries. The possibility of isolating a gravitational contribution to the stress-energy tensor therefore does not amount to identifying an autonomous gravitational energy, and neither reading commits us to gravitational energy as an autonomous entity in its own right. However, we conclude that gravitational (stress-) energy is a meaningful concept nonetheless and supports gravity as a matter-type field. 

A word on scope is in order. Throughout we treat gravity as a \textit{classical} field: the metric $g_{\mu\nu}$ remains a c-number background and only non-gravitational fields are quantized. This is the setting of quantum field theory on a curved spacetime, or semiclassical gravity, in which the classical metric is sourced by the expectation value of the renormalized quantum stress-energy tensor through the semiclassical Einstein equations. We make no attempt to quantize the gravitational field itself, and none of our conclusions rests on a theory of quantum gravity. We take this restriction to be a virtue rather than a limitation. It is a deliberately \textit{humble} step: instead of adjudicating the status of matter and spacetime from within one or another still-speculative approach to quantum gravity, we ask what already befalls the matter/spacetime distinction at the comparatively well-charted semiclassical level—arguably the first place at which quantum theory and gravitation are forced to share a single stress-energy tensor. That the distinction should come under pressure already here is precisely what we find of interest\footnote{The semiclassical description has a limited regime of validity: it is expected to break down once fluctuations of the stress-energy become large, where a stochastic or fully quantum-gravitational treatment would be required.  In any case, it seems highly unlikely that the outcome will differ in those cases, as the presence of renormalized stress-energy tensors is still required in extended versions of semiclassical gravity \citep{Hu:2008rga}.}.

The paper is organized as follows. In §\ref{sec:classical} we recall the classical criterion for matter of \cite{Ferreiro:2025fii}: it qualifies the non-gravitational fields as matter and excludes the metric. In §\ref{sec:quantum} we pass to quantum field theory on a classical curved spacetime: renormalizing the stress-energy operator forces a local, purely geometric term into the tensor, the non-gravitational fields keep their material status, and both classical arguments fail---so a stress-energy contribution for the metric can, after all, be written. In §\ref{sec:egal-unif} we draw the two readings this makes available: the \textit{egalitarian} view (§\ref{sec:egalitarian}), on which renormalization gives the metric a material significance alongside its spatio-temporal one, and the \textit{unificationist} view (§\ref{sec:unificationist}), on which gravitational and non-gravitational energy cannot be split without a choice of vacuum a generic spacetime does not supply. Together they pose a dilemma---either both energies are meaningless, leaving only classical fields as matter, or both are meaningful, and the metric is itself a matter field---on neither horn of which the classical matter/spacetime divide survives quantization. Finally, §\ref{sec:conclusions} gathers the conclusions and further open problems raised by combining quantum field theory with classical gravity.

\section{Matter in Classical Field Theory}
\label{sec:classical}
%In this section we introduce the main criterion we use for calling a set of fields matter fields. As discussed in the introduction the advantage of this criterion is (i) it includes all paradigmatic matter fields, e.g. electromagnetic, fluids, scalar fields, etc., and (ii) it excludes the metric $g_{\mu\nu}$ of General Relativity.

Two essential properties we ascribe to matter are (i) the existence of a dynamics in terms of equations of motion and (ii) the existence of a stress-energy tensor. As an example take a scalar field $\phi$ coupled to a current $J(x)$ with the following action
\begin{align}
    S[\phi, J, g_{\mu\nu}]=-\frac12\int d^4x \sqrt{-g(x)}\left(g^{\mu\nu}(x)\nabla_\mu\phi(x)\nabla_\nu\phi(x)+ \alpha\phi^2(x) J(x) \right).
    \label{eq.action}
\end{align}
The first term is just the kinetic term of the scalar field $\phi$ and the second term is an interaction term between $\phi$ and $J$ mediated by the coupling constant $\alpha$. The associated equations of motion and stress-energy tensor are:
\begin{align}
  & \nabla_\mu \nabla^\mu \phi(x)= \alpha J(x) \phi(x),  \\&T_{\mu\nu}^\phi(x)=\nabla_\mu\phi(x) \nabla_\nu \phi(x)-\frac12 g_{\mu\nu}\nabla^\alpha\phi(x) \nabla_\alpha \phi(x)-\frac{\alpha}{2}g_{\mu\nu}(x)J(x)\phi^2(x).
   \label{eq.classical.stress.energy}
\end{align}
 It is easy to check that on-shell (when the equations of motion are satisfied) the stress-energy tensor obeys
\begin{align}
\nabla^\mu T_{\mu\nu}^\phi(x)\underbrace{=}_{\text {on-shell}}-\frac{\alpha}{2}\phi(x)^2\nabla_\nu J(x).
\label{eq.classical.conservation}
\end{align}
At this point we could say that $\phi$ is indeed a matter field. The next question we can ask is what is $J(x)$. Physically we can think of it as an external source, in analogy to an electromagnetic current in the Maxwell equations. In order to determine if it is a matter field we would also require it to have a dynamics and a stress-energy tensor. We can choose for instance to add to the original action a kinetic term for $J(x)$ which would imply the associated equations of motion and stress-energy tensor
\begin{align}
   \nabla_\mu \nabla^\mu J(x)= \frac{\alpha}{2}\phi^2(x), \quad \quad T_{\mu\nu}^J(x)=\nabla_\mu J(x) \nabla_\nu J(x)-\frac12 g_{\mu\nu}(x)\nabla^\alpha J(x) \nabla_\alpha J(x).
\end{align}
%Now, given the two matter fields $\phi(x)$ and $J(x)$ we observe two things. First, $J$ couples to $\phi$ in the sense that it appears in the dynamical equation of $\phi$ and viceversa, $\phi$ couples to $J(x)$. We can identify this as the action-reaction principle that we associated with matter. We say in this case that $\phi$ and $J$ are \textit{directly coupling}. One could \rev{generalize}{generalise} this to $n$-tuple of fields but with an important clarification. Take a field $\psi(x)$ such that $J$ and $\psi$ are directly coupled, but it does not appear in the dynamics of $\phi$ and viceversa, i.e., it is \textit{directly un-coupled}. Since $\phi$ and $J$ are directly coupled, we say a $n$-tuple of fields are \textit{C-interacting} if each of them is directly coupled to at least one other field of the tuple.
Then, the full stress-energy tensor $T_{\mu\nu}^\phi(x)+T_{\mu\nu}^J(x)$ now yields
\begin{equation}
    \nabla^\mu \left(T_{\mu\nu}^\phi(x)+T_{\mu\nu}^J(x)\right)\underbrace{=}_{\text {on-shell}}0 \label{conservT}
\end{equation}
The lesson we can extract from this case is the following. Matter in the form of fields has an associated dynamics and (stress-)energy and, in the interacting case, an associated (stress-)energy exchange. With this in mind we propose the minimal criterion for matter:
%We interpret this condition as the fact that the subsystems is \textit{E-interacting}, i.e. there is only one conserved stress-energy associated to both fields. With this in mind we propose the following criterion for a system of matter $\phi_1,\phi_2,...,\phi_n$ to be matter:
\begin{criteria}[Minimal\footnote{For a more elaborated criterion see \cite{Ferreiro:2025fii}.} Classical Matter criterion]\label{crit:1}
Given a theory $\mathcal{T}$, composed of relativistic classical fields, we say the fields $M=\left\{\phi_1,\phi_2,...,\phi_n\right\}$ are all matter fields if for any given physical setup of $M$:
\begin{enumerate}
    \item There exist dynamical equations for each field in $S$ and a stress-energy $T^M_{\mu\nu}$ that is composed of all fields in $S$.
    \item  If the fields are interacting dynamically, then they interact via (stress-)energy exchange.
    %\item If $\nabla_\mu T_S^{\mu\nu}=0$, we can partition the system $S$ into E-interacting subsystems $S_i$ iff the system can be reduced to a sum of irreducible stress-energy tensors each conserved independently $\nabla_\mu T_{S_i}^{\mu\nu}=0$.
    %\item Each E-interacting subsystems is a C-interacting subsystem (and viceversa).
\end{enumerate}
\end{criteria}
%This criterion looks abstract but the underlying reasoning is the if a matter system $S$ can be further divided into subsystems $S_i$ that don't couple to each other via its dynamical equations then there is no transference of energy between them.
All known (non-gravitational) fields that we usually attribute to matter, e.g., electromagnetic fields and fluids, satisfy Criterion \ref{crit:1}. In the following we analyze the case of the gravitational field, more specifically the metric $g_{\mu\nu}$ in General Relativity. 
\subsection{Gravitational (stress-) energy in General Relativity}
Given the above-mentioned criterion for matter it is natural to ask: is the gravitational field in General Relativity, namely the metric $g_{\mu\nu}$, also a matter field? Two arguments count against it. The first turns on the exchange of energy. Naively, the metric has dynamical equations, i.e., the Einstein equations
\begin{align}
    R_{\mu\nu}(x)-\frac12 g_{\mu\nu}(x)R(x)= 8\pi G \  T_{\mu\nu}(x), \label{EEQ}
\end{align}
and in the presence of any matter field beyond $g_{\mu\nu}$ the stress-energy tensor will also include the metric $g_{\mu\nu}$. The deciding point is condition 2 of criterion \ref{crit:1}. If $g_{\mu\nu}$ were a matter field, it would interact dynamically with all matter fields, which is its \textit{universality} property. However, there is no way of constructing an analogue of \eqref{conservT} for the metric, since by default any field that only interacts with the metric is conserved \footnote{The standing debate rests on the possible interpretations in terms of pseudo-tensors \citep{read2018, duerr2019a, duerr2019b}.}.

The second argument concerns the very existence of a gravitational stress-energy. Even if one rejects condition 2, the elephant in the room remains: could there be such a thing as gravitational stress-energy? Could one construct a candidate?  To our knowledge, the best argument against a stress-energy \textit{tensor} associated to gravity was developed in \cite{Curiel:2018gfn}. We briefly reconstruct the essential steps of this argument, recast in the
language of the present section. The strategy is to lay down the conditions
that any such object would necessarily have to meet, and then to prove that no
tensor can meet all of them at once.

The first step is to fix what it would even mean for a tensor to be the
stress-energy of the metric. Just as $T_{\mu\nu}^\phi$ is a definite function
of $\phi$ and its derivatives, i.e., the same functional expression in every
spacetime, irrespective of the particular solution, a gravitational
stress-energy tensor would have to be a definite function of the metric and its
derivatives, computable wherever there are gravitational phenomena. Following
\cite{Curiel:2018gfn} we call such a systematically metric-dependent object a
\emph{concomitant} of $g_{\mu\nu}$. Since the connection can be
made to vanish along any given curve by a choice of coordinates, no genuine
energetic quantity can depend on it: the object must depend on the metric only
through the curvature, i.e. it must in fact be a concomitant of the Riemann
tensor $R^{\mu}{}_{\nu\alpha\beta}$.

The remaining conditions are read off from the very properties that make
$T_{\mu\nu}^\phi$ and $T_{\mu\nu}^J$ above legitimate stress-energy tensors.
Being a stress-energy tensor, the candidate $T_{\mu\nu}^{\rm Grav}$ must be a two-index, symmetric tensor. It must be covariantly divergence-free whenever the
stress-energy of the ponderable matter vanishes, in the sense of the
conservation law we associated above \eqref{conservT}. It must not
vanish wherever the curvature is non-trivial, for one can always envisage a
device that extracts energy from tidal forces, i.e. from geodesic deviation.
And, finally, it must carry the physical dimension of stress-energy, for this is
precisely what allowed us to form the sum $T_{\mu\nu}^\phi+T_{\mu\nu}^J$ and
speak of a transference of energy between the subsystems. To make this last
requirement precise we appeal to the Einstein equation \eqref{EEQ} itself: demanding that
\eqref{EEQ} still hold after a constant
rescaling $g_{\mu\nu}\to\lambda\, g_{\mu\nu}$, and noting that in geometrized
units $G_{\mu\nu}$ (with both indices lowered) is invariant under such a rescaling while Newton's constant
carries a fixed dimension, we conclude that any $T_{\mu\nu}$ standing on the
right-hand side must likewise be invariant. Hence carrying the dimension of
stress-energy amounts to demanding that $T_{\mu\nu}^{\rm Grav}$ be a homogeneous
concomitant of weight zero. We collect these requirements in the following
theorem \citep{Curiel:2018gfn}.
\begin{theorem}[Non-existence of a gravitational stress-energy tensor]
\label{thm:nogse}
There is no tensor field $T^{\mu\nu}_{\rm Grav}$ satisfying simultaneously the
following conditions:
\begin{enumerate}
    \item $T_{\mu\nu}^{\rm Grav}$ is a concomitant of the
    Riemann tensor $R^{\mu}{}_{\nu\alpha\beta}$, that is, a systematic,
    coordinate-independent function of the curvature, the same in every
    spacetime;
    \item $T_{\mu\nu}^{\rm Grav}$ is a two covariant-index,
    symmetric tensor;
    \item $T_{\mu\nu}^{\rm Grav}$ has the physical dimension
    of stress-energy, i.e. it is homogeneous of weight zero, being invariant
    under a constant rescaling $g_{\mu\nu}\to\lambda\, g_{\mu\nu}$;
    \item  $T_{\mu\nu}^{\rm Grav}$ is covariantly
    divergence-free, $\nabla^\mu T_{\mu\nu}^{\rm Grav}=0$, whenever the
    stress-energy of the ponderable matter vanishes;
    \item  $T_{\mu\nu}^{\rm Grav}\neq 0$ at every point
    at which $R^{\mu}{}_{\nu\alpha\beta}\neq 0$.
\end{enumerate}
\end{theorem}

Here we present a sketch of the proof. Conditions (1) -- (3) already fix the differential order of the candidate. A
homogeneous concomitant of the metric of order $n$ scales as $\lambda^{1-n/2}$
under $g_{\mu\nu}\to\lambda\, g_{\mu\nu}$, so weight zero forces $n=2$: the
object is built algebraically from the metric and its first two derivatives, or
equivalently it is a zeroth-order concomitant of the Riemann tensor. But the
only two covariant-index, zeroth-order concomitants of the Riemann tensor are
the linear combinations
\begin{equation}
    T_{\mu\nu}^{\rm Grav}=a\,R_{\mu\nu}+b\,R\,g_{\mu\nu}, \qquad a,b\in\mathbb{R}.
\end{equation}
Imposing condition (4), the only such combinations with vanishing covariant
divergence are the constant multiples of the Einstein tensor,
$T_{\mu\nu}^{\rm Grav}=k\,G_{\mu\nu}$ with $G_{\mu\nu}=R_{\mu\nu}-\tfrac12
R\,g_{\mu\nu}$. This exhausts the candidates. Yet $G_{\mu\nu}$ vanishes exactly
where $R_{\mu\nu}$ vanishes, and there exist spacetimes with $R_{\mu\nu}=0$ but
$R^{\mu}{}_{\nu\alpha\beta}\neq 0$---any vacuum region with non-vanishing Weyl
curvature, such as the exterior Schwarzschild solution or a gravitational
wave. At such points the only surviving candidate is zero while the curvature
is not, in direct contradiction with condition (5). Hence no tensor satisfies
(1)--(5) simultaneously.

 The moral, in the
terms of our Criterion~\ref{crit:1}, is immediate:
if $g_{\mu\nu}$ were part of the matter content
$M$ for physical systems $S$ where all fields except $g_{\mu\nu}$ vanish, i.e., vacuum solutions, there would be no stress-energy, since the only tensor the metric supplies
through its curvature that could play that role is a multiple of $G_{\mu\nu}$, and in this system it is just zero. The metric is therefore not a matter field. This would imply that entities such as static black holes or gravitational waves do not carry energy\footnote{A possible way out of this is of course to construct non-local objects (see \cite{Curiel:2018gfn}).}.
%\paragraph{Argument 2}. Take now the system of fields used in §2.1  $S=\{\phi(x),J(x)\}$ but they are not interacting, i.e., $\alpha=0$. Now suppose we would include the metric $g_{\mu\nu}(x)$ in the subsystem. Then from criterion 1, (i) is satisfied and so is (ii), $g_{\mu\nu}(x)$ is directly coupled to $\phi(x)$ and $J(x)$, and these two are directly un-coupled. For condition (iii) one can partition the system into E-interacting subsystems $S_1$ with (conserved) stress-energy tensor $T^{\mu\nu}_\phi$ and another with $T^{\mu\nu}_J$. However, in terms of C-interacting system there is only one system  $S'=\{\phi(x),J(x),g_{\mu\nu}(x)\}$. In order words, the three fields are interacting at the level of the equations of motions but not at the level of energy transference, therefore condition (iv) is not satisfied and indeed $g_{\mu\nu}$ is not a matter field.
\section{Matter in Quantum Field Theory}
\label{sec:quantum}

In this section we reassess the discussion of the previous section for quantum field theories. As anticipated in the introduction, we take such a theory to rest on three ingredients: a classical field theory given by the action \eqref{eq.action}; a \textit{quantization}, under which the classical fields become operators $\phi \to \hat{\phi}$ (we maintain $J$ as a classical c-field); and a state, which assigns expectation values to the resulting observables. Our attention will rest mainly on the second ingredient: because $\hat{\phi}$ is an operator the stress-energy tensor, being a product of fields at a point, is no longer well defined, and recovering it requires a renormalization that introduces new terms with no classical counterpart \citep{Wald:1995yp}.

The algebraic approach provides a local and covariant formulation of the QFT associated with the action~\eqref{eq.action}. Quantum fields $\hat\phi$ are defined as algebra-valued distributions over $C^\infty_0(M)$; the smeared field is
\begin{equation}
  \hat\phi(f)=\int_M d^4x\,\sqrt{-g}\,f(x)\,\hat\phi(x),
\end{equation} and the algebra of observables $\A$ is generated by the $\hat\phi(f)$ subject to linearity, the field equation $\hat\phi(Pf)=0$ with $P=\Box-\alpha J$, hermiticity, and the commutator $[\hat\phi(f_1),\hat\phi(f_2)]=iE(f_1,f_2)\,\unit$, with $E$ the causal propagator. A state is a normalized positive functional
$\omega:\A\to\mathbb C$; its expectation values $\langle A\rangle_\omega =\omega(A)$ are recovered, via GNS, on a Hilbert space \citep{Hollands:2014eia}.

Before turning to the stress-energy tensor, we observe that the dynamical part of Criterion~\ref{crit:1}, i.e., the requirement that each field obey an equation of motion, carries over to the quantum theory unchanged. In the theory considered here the field equation is the operator relation
$\hat\phi(Pf)=0$, the smeared and quantized form of the classical
Euler--Lagrange equation, with the coupling to the source encoded in
$P=\Box-\alpha J$ exactly as in the classical case.

The stress-energy tensor requires products of fields at a point, which are
ill defined in $\A$ as it stands. The remedy is to enlarge $\A$ to its algebra of Wick polynomials, in which local quadratic observables acquire a well-defined meaning as operator-valued distributions
\citep{Hollands:2001nf,Moretti:2001qh,Hollands:2004yh}. Concretely, one
point-splits the classical tensor and defines
\begin{equation}
  \mathrm{T}_{\mu\nu}(x,x')
   \equiv g_\nu{}^{\nu'}\nabla_\mu\nabla_{\nu'}
     -\tfrac12 g_{\mu\nu}g^{\rho\alpha'}\nabla_\rho\nabla_{\alpha'}
     -\tfrac{\alpha}{2}g_{\mu\nu}J ,
\end{equation}
where $\nu'$ denotes derivative with respect to $x'$. Applied to $\phi(x)\phi(x')$ and taking the coincident limit recovers \eqref{eq.classical.stress.energy}. However, applied to $\hat\phi(x)\hat\phi(x')$ the coincident limit diverges, but this divergence can be isolated and subtracted. This is exactly what is encoded in the so-called \emph{Hadamard parametrix} $H(x,x')$ -- a local,
state-independent singular bidistribution fixed by the geometry and $P$
alone, such that we can construct:
\begin{equation}\label{eq:Twick}
  \lim_{x'\to x}\mathrm{T}_{\mu\nu}(x,x')
     \bigl[\hat\phi(x)\hat\phi(x')-H(x,x')\,\unit\bigr],
\end{equation}
which is indeed well defined. Because $H$ is subtracted rather than a state's two-point function, this piece is a genuine element of the (extended) algebra: an \emph{operator}, defined with no reference to $\omega$.
This operator is, however, not conserved: the subtracted parametrix is
only a bisolution of the field equation up to a smooth remainder, so the
classical route to $\nabla^\mu T_{\mu\nu}=0$ (which relied on going
on-shell) is no longer automatic \citep{Moretti:2001qh,Hollands:2004yh}.
Conservation is restored at the operator level by defining
\begin{equation}\label{eq:improve}
  \wick{\hat T_{\mu\nu}}(x)
  \equiv \lim_{x'\to x}\mathrm{T}_{\mu\nu}(x,x')
     \bigl[\hat\phi(x)\hat\phi(x')-H(x,x')\,\unit\bigr]+C_{\mu\nu}(x)\unit
\end{equation}
with $C_{\mu\nu}(x)$ the local, background-built tensor
\begin{equation}\label{eq:Cterm}
  C_{\mu\nu}(x)=\frac{g_{\mu\nu}}{16\pi^2}\Bigl(
    \tfrac{1}{180}R_{\alpha\beta\gamma\delta}^2
   -\tfrac{1}{180}R_{\alpha\beta}^2
   +\tfrac{1}{30}\Box R
   +\tfrac{1}{72}R^2
   +\tfrac{\alpha^2}{2}J^2
   -\tfrac{\alpha}{6}\Box J
   -\tfrac{\alpha}{6}RJ\Bigr),
\end{equation}
such that
\begin{equation}
   \nabla^\mu \wick{\hat T_{\mu\nu}}(x)=-\frac{\alpha}{2}\nabla_\nu J(x)\wick{\hat \phi^2}(x) \label{eq.quantum.conservation}
\end{equation}
which is exactly the quantum version of
\eqref{eq.classical.conservation}. $\wick{\hat T_{\mu\nu}}$ is a genuine operator, fixed uniquely up to admissible local curvature
terms $\Theta_{\mu\nu}$ (the conserved combinations\footnote{These two terms are obtained by varying $\int\sqrt{-g}\,R^2$ and $\int\sqrt{-g}\,R_{\mu\nu}R^{\mu\nu}$ with respect to the metric.} $^{(1)}\!H_{\mu\nu}$,
$^{(2)}\!H_{\mu\nu}$, together with $G_{\mu\nu}$ and $g_{\mu\nu}$),  all of them multiples of $\unit$. Promoting $\phi$ to a quantum field therefore changes the structure
of the stress-energy tensor as an observable: to make it well defined we must introduce the function $H(x,x')$ and to make it compatible with the classical conservation law one
must include a local, background-dependent contribution $C_{\mu\nu}(x)$,
present in the operator before any state is chosen.

In analogy with §\ref{sec:classical} we can use the dynamics and the stress-energy tensor associated with $J$ to find a fully conserved stress-energy tensor. Of course, finding a stress-energy tensor whose sum with $\wick{\hat T_{\mu\nu}}$ is divergence-free requires an operator in the algebra. A good candidate would be the stress-energy tensor of $J$ itself. Since $J$ is
a classical, c-number field, its stress-energy tensor
\begin{equation}\label{eq.TJ.classical}
  T^J_{\mu\nu}=\nabla_\mu J\,\nabla_\nu J
    -\tfrac12\,g_{\mu\nu}\,\nabla^\rho J\,\nabla_\rho J
\end{equation}
carries no ordering ambiguity and requires no renormalization: it is already a
number at each point. To place it in $\A$ on the same footing as
$\wick{\hat T_{\mu\nu}}$ we identify it with a multiple of the identity,
\begin{equation}\label{eq.TJ.operator}
  \wick{\hat T^J_{\mu\nu}}(x)\equiv T^J_{\mu\nu}(x)\,\unit ,
\end{equation}
exactly as the counterterm $C_{\mu\nu}(x)$ entered \eqref{eq:improve}. Adding \eqref{eq.TJ.operator} to $\wick{\hat T_{\mu\nu}}$ does not yet yield a
conserved object. Using \eqref{eq.quantum.conservation} together with
$\nabla^\mu T^J_{\mu\nu}=(\Box J)\,\nabla_\nu J$, the divergence of the sum is
\begin{equation}\label{eq.sum.divergence}
  \nabla^\mu\bigl(\wick{\hat T_{\mu\nu}}+T^J_{\mu\nu}\,\unit\bigr)
   =-\frac{\alpha}{2}\,\nabla_\nu J\,\wick{\hat\phi^2}
    +(\Box J)\,\nabla_\nu J\,\unit .
\end{equation}
The two terms cannot cancel as operators: the first is proportional to
$\wick{\hat\phi^2}$, a genuine element of the algebra, while the second is a
c-number multiple of $\unit$. No further operator built from $J$ alone can
supply the missing balance, precisely because $J$ enters only through $\unit$.
The counterterm $C_{\mu\nu}(x)$ has brought the divergence of
$\wick{\hat T_{\mu\nu}}$ into agreement with the classical exchange law
\eqref{eq.quantum.conservation}, but it has not, by itself, made the
stress-energy tensor conserved.

The obstruction is removed once we pass to expectation values. Taking the state
$\omega$ in \eqref{eq.sum.divergence}, the operator $\wick{\hat\phi^2}$ is
replaced by the c-number $\langle\wick{\hat\phi^2}\rangle_\omega$, and
\begin{equation}\label{eq.exp.divergence}
  \nabla^\mu\bigl\langle\,\wick{\hat T_{\mu\nu}}+T^J_{\mu\nu}\,\unit\,\bigr\rangle_\omega
   =\Bigl(\Box J-\frac{\alpha}{2}\,\langle\wick{\hat\phi^2}\rangle_\omega\Bigr)\nabla_\nu J .
\end{equation}
This vanishes provided $J$ obeys its own equation of motion, now
sourced by the expectation value of the quantum field,
\begin{equation}\label{eq.Jeom}
  \Box J=\frac{\alpha}{2}\,\langle\wick{\hat\phi^2}\rangle_\omega ,
\end{equation}
whereupon
\begin{equation}\label{eq.total.conservation}
  \nabla^\mu\bigl\langle\,\wick{\hat T_{\mu\nu}}+T^J_{\mu\nu}\,\unit\,\bigr\rangle_\omega=0 .
\end{equation}

Two ingredients are thus required to recover a conserved stress-energy tensor. The counterterm $C_{\mu\nu}(x)$ is needed already
at the operator level, to align the divergence of $\wick{\hat T_{\mu\nu}}$ with
the classical exchange law; it does not render the tensor conserved. Conservation
follows only after the second step: closing the system with the stress-energy of
$J$ and its dynamical equation \eqref{eq.Jeom}. Because the source of that
equation is an expectation value, the closure---and with it the conservation law
\eqref{eq.total.conservation} -- can be imposed only on expectation values, not as
an identity between operators. This is the semiclassical, mean-field character of
the construction: the sharp classical field $J$ is driven by the quantum mean
$\langle\wick{\hat\phi^2}\rangle_\omega$, and it is at this level, and only this
level, that the energy exchanged between the two sectors balances.

The construction of the previous paragraphs isolates precisely what survives
quantization and what is modified, and this is what we now use to adapt
Criterion~\ref{crit:1} to the quantum theory. Two lessons are relevant. First,
the dynamical content of the classical criterion -- the existence of equations of
motion -- is inherited
unchanged, now at the level of operator field equations. Second, the energetic
content is inherited only in a weaker form: the stress-energy tensor of a
subsystem is a renormalized operator, whose field-dependent part is built from
the fields of the subsystem but which also carries a local, background-built
counterterm proportional to $\unit$, and whose conservation is in general not an
operator identity but a statement about expectation values, holding once the
semiclassical equations of motion are posed. A naive inheritance,
declaring a quantum tuple to be matter whenever its classical counterpart
satisfies Criterion~\ref{crit:1}, would record the first lesson but miss the
second. We adapt the criterion so that both are built in.
\begin{criteria}[Minimal Quantum Matter criterion]\label{crit:2}
Given a theory $\mathcal{T}$, composed of relativistic quantum and classical fields, we say the fields $M=\{\hat\phi_1,\dots,\hat\phi_m;\,J_1,\dots,J_k\}$ are all matter fields if for any given physical system $S$:
\begin{enumerate}
    \item There exist dynamical equations for each field in $S$ and a stress-energy operator $\wick{\hat{T}_M^{\mu\nu}}$ that is composed of all fields in $S$.
    \item  If the fields are interacting dynamically, then they interact via exchange of the expectation value of (stress-)energy.
\end{enumerate}
\end{criteria}

Criterion~\ref{crit:2} reduces to Criterion~\ref{crit:1} in the classical limit,
where the counterterm and the ordering ambiguity vanish and expectation values
collapse to their classical values. The tuple $\{\hat\phi,J\}$ satisfies it: both
fields obey equations of motion, have a stress-energy tensor
$\wick{\hat T_{\mu\nu}}+T^{J}_{\mu\nu}\unit$, and interchange energy via the expectation value \eqref{eq.exp.divergence}.
\subsection{Stress-energy for gravity in semiclassical Einstein theory}
Now we can re-ask the question: can the metric field $g_{\mu\nu}$ be a matter field in light of Criterion~\ref{crit:2}? Again, from the side of the dynamical equations no obstacle appears on the QFT side as long as we provide a dynamical equation for $g_{\mu\nu}$. The canonical way of doing so is via the \textit{semiclassical Einstein equation}
\begin{align}
    R_{\mu\nu}(x)-\frac12 g_{\mu\nu}(x)R(x)= 8\pi G \ \bigl\langle\,\wick{\hat T_{\mu\nu}}\bigr\rangle_\omega. \label{SEEQ}
\end{align}

Looking now at condition 2 and \eqref{eq:improve}, we realize that there is a well-suited candidate for a gravitational stress-energy tensor, namely\footnote{We take here $J=0$ for simplicity but this does not affect the conclusions.}
\begin{equation}\label{eq:Tgrav}
  T_{\mu\nu}^{\rm Grav} \equiv  C_{\mu\nu}(x)+c_0\, g_{\mu\nu}+c_1\, G_{\mu\nu}+c_2\,^{(1)} H_{\mu\nu}+c_3\, ^{(2)}H_{\mu\nu}
\end{equation}
such that
\begin{equation}\label{eq:grav-conservation}
  \nabla^\mu\wick{\hat T_{\mu\nu}}(x)
  = \nabla^{\mu}\left(\lim_{x'\to x}\mathrm{T}_{\mu\nu}(x,x')
     \bigl[\hat\phi(x)\hat\phi(x')-H(x,x')\,\unit\bigr]+ T_{\mu\nu}^{\rm Grav}\unit\right)=0 .
\end{equation}

The argument for the non-existence of a gravitational stress-energy tensor rested on Theorem~\ref{thm:nogse}, the non-existence of a tensor
meeting all five conditions on a gravitational stress-energy. These conditions
do not carry over to the semiclassical theory. The term $T_{\mu\nu}^{\rm Grav}$ meets the first
two conditions\footnote{Provided $c_0=0$.} -- it is a local, covariant, symmetric rank-two concomitant of
the metric -- but not the last three, and for different reasons.

On condition (3) of Theorem 1, the point-splitting prescription is not scale invariant:
under $g_{\mu\nu}\to\lambda\,g_{\mu\nu}$ the renormalized tensor acquires a term
proportional to $\ln\lambda$ and to a conserved local curvature tensor, so the
weight-zero condition (3) already fails. The failure lies in the inapplicability of the assumption from
which condition (3) is derived. That assumption is that the candidate
sits on the right-hand side of an Einstein equation, opposite the weight-zero
$G_{\mu\nu}$, and must therefore itself be weight zero. In the semiclassical
theory that very equation is not scale invariant.

This is seen most sharply for a conformally coupled massless scalar\footnote{\label{fn:conformal}The construction of the stress-energy tensor has been shown for simplicity for the minimally coupled scalar field. The conformally coupled scalar fields can be seen as a type of scalar--tensor theory, which is the focus of \cite{Ferreiro:2025fii}. However, in our case we can regard the conformally coupled scalar case as a representation of other conformally invariant fields that do not couple to the curvature, e.g. the electromagnetic field or the massless Dirac field.}. Classically
such a field is Weyl invariant and its stress-energy tensor is traceless
on-shell, $T^{\mu}{}_{\mu}=0$; the trace of the Einstein equation then reads
$-R=0$, which is preserved under $g_{\mu\nu}\to\lambda\,g_{\mu\nu}$, both sides
being simply zero. Quantization breaks this classical Weyl invariance. The
renormalized trace no longer vanishes but equals the \emph{conformal (trace)
anomaly}, a purely geometric, state-independent multiple of the identity built
from curvature invariants of mass dimension four,
\begin{equation}\label{eq:anomaly}
  \langle {:}\hat T^{\mu}{}_{\mu}{:}\rangle_{\omega}
   = c_{_W}\,C_{\alpha\beta\gamma\delta}C^{\alpha\beta\gamma\delta}
     - a_{_E}\,E_{4} + b\,\Box R ,
\end{equation}
with $C_{\alpha\beta\gamma\delta}$ the Weyl tensor and $E_{4}$ the Euler density.
The central charges are fixed numbers, independent of the state
($c_{_W}=1/1920\pi^{2}$ and $a_{_E}=1/5760\pi^{2}$ for a single scalar); only the
coefficient $b$ of the $\Box R$ term is renormalization-scheme dependent, and can
be shifted by the ambiguity $\Theta_{\mu\nu}$. Decisively, the anomaly is the
same in every Hadamard state: it is not an artefact of a particular choice of
$\omega$ but a fixed feature of the geometry.

Taking the trace of the semiclassical Einstein equation \eqref{SEEQ} now gives
\begin{equation}\label{eq:trace-sce}
  -R \;=\; \frac{8\pi G}{c^{4}}\,
           \langle {:}\hat T^{\mu}{}_{\mu}{:}\rangle_{\omega}
        \;=\; 8\pi G \
           \bigl( c_{_W}\,C_{\alpha\beta\gamma\delta}C^{\alpha\beta\gamma\delta}
                  - a_{_E}\,E_{4} + b\,\Box R \bigr) .
\end{equation}
Under the constant rescaling $g_{\mu\nu}\to\lambda\,g_{\mu\nu}$ the two sides
transform with \emph{different} weights: the scalar curvature scales as
$R\to\lambda^{-1}R$, whereas every term on the right, being quadratic in the
curvature, scales as $\lambda^{-2}$. No single weight can be assigned to
\eqref{eq:trace-sce} consistently; the anomaly obstructs its invariance under
rescaling. And because the trace is exactly the state-independent part of
$\langle {:}\hat T_{\mu\nu}{:}\rangle_{\omega}$, this obstruction can be removed
neither by a choice of state nor by the scheme freedom $\Theta_{\mu\nu}$, which
reaches only the $\Box R$ term. The premise underlying Curiel's dimension
condition---a semiclassical Einstein equation invariant under
$g_{\mu\nu}\to\lambda\,g_{\mu\nu}$, which is what would force its source to be
weight zero---therefore does not hold once the matter is quantized. Condition (3)
fails because the scale invariance
that would have demanded it is itself anomalously broken.

It is not separately conserved, but this is a demand on a complete and autonomous  gravitational
stress-energy tensor, i.e., whenever the stress-energy tensor of the ponderable matter vanishes. However, it is not clear what this means in the quantum field theoretic case, and we will return to this in §\ref{sec:unificationist}. Vacuum-type solutions, e.g. black holes or emitting gravitational waves from black hole binaries will in general have non-vanishing $C_{\mu\nu}(x)$, but it could be zero in very specific and idealized cases (see §\ref{sec:conclusions}).

Under this view, $ T_{\mu\nu}^{\rm Grav}$ is an acceptable candidate for a gravitational stress-energy tensor and nothing in Criterion~\ref{crit:2} contradicts the assessment of $g_{\mu\nu}$ as a matter field\footnote{Note that this is only valid for quantum fields, not classical fields. The $J$ fields interact dynamically with $g_{\mu\nu}$ but they do not exchange energy with it.}. This opens the question: if $ T^{\mu\nu}_{\rm Grav}$ is an acceptable candidate for a gravitational stress-energy tensor, is it sufficient to declare gravity a matter field? In §\ref{sec:egal-unif} we develop two readings of this result. The first is the \textit{egalitarian} view, which ascribes to $g_{\mu\nu}$ both spatio-temporal and material significance. The second is the \textit{unificationist} view, a deflationary position on the split between gravitational and non-gravitational energy, grounded in the property of gravity that sets it apart from all other fields---its universality. Neither grants gravitational energy an autonomous existence but supports a matter behavior for $g_{\mu\nu}$.

\section{An egalitarian and unificationist view on gravity and matter}
\label{sec:egal-unif}

In this section we defend two positions on the relation between gravity and matter, taking into account both General Relativity and Quantum Field Theory: a (weak) egalitarian view of the metric $g_{\mu\nu}$, i.e. that it has both gravitational and material significance, and a unificationist view on
gravitational and non-gravitational energy. The two are logically independent:
the first requires only that the renormalization term $C_{\mu\nu}(x)$ be a
genuine energetic contribution built from the metric; the second, the stronger
claim, requires in addition that the division of the total energetic content
into a matter part and a gravitational part not be fixed by the theory. 

\subsection{Egalitarian view on spacetime and gravity (as a matter field)}
\label{sec:egalitarian}

One of the central interpretive questions raised by the theory of General
Relativity (GR) concerns the relationship between gravitation and the geometry
of spacetime. \cite{LEHMKUHL200883} distinguishes three families of positions on this
question. On the geometric interpretation, gravity is reducible to, or is
nothing more than, a manifestation of spacetime geometry; on the field
interpretation, conversely, spacetime geometry is reducible to, or a
manifestation of, the behavior of the gravitational field; and on the
egalitarian interpretation, gravity and spacetime geometry are neither reducible to one another, but are in some sense identified. Whereas the first two are
reductive positions, each taking one of the two notions to be more fundamental
than the other, egalitarianism treats them as standing on a par.

Egalitarianism, \cite{LEHMKUHL200883}  argues, is itself a family rather than a single thesis,
admitting of at least three grades. According to weak egalitarianism, every
mathematical object in the formalism of GR -- the metric $g_{ab}$, the affine
connection $\Gamma$, and the Riemann curvature tensor alike -- possesses both
geometrical and gravitational significance, playing a role in the description
of both geometrical phenomena (durations, lengths, angles, the paths of freely
falling bodies) and gravitational phenomena. Moderate egalitarianism strengthens this into a claim of one-to-one correspondence, on which any gravitational description of a phenomenon may be reformulated as a geometrical one and vice versa. Strong egalitarianism goes further still,
holding that gravity and geometry are not merely correlated but conceptually
identified. \cite{LEHMKUHL200883}  maintains that the weak version is enforced by the formalism, since every object of the theory does carry both kinds of significance.

We support a weak egalitarian view of the metric, proposing an addition that classical GR does not supply. In \cite{LEHMKUHL200883}, the two axes are the geometrical and the gravitational; the quantum treatment forces a third, which we call \emph{material} or \emph{energetic} significance: a mathematical object has material significance if it enters the stress-energy content of the theory, that is, if it contributes to the conserved energetic balance. Classically, the metric lacks this significance, but §\ref{sec:quantum} established that it acquires it at the semiclassical level through the purely geometric term $C_{\mu\nu}(x)$, introduced by renormalization and required for conservation. Our claim, then, is that in quantum field theory on a curved background the metric is not only a geometrical and a gravitational object, as in \cite{LEHMKUHL200883}, but also a material one, not merely a field, but a matter field.
%\andrea{does the material significance replace gravitational? Or we cannot say at semiclassical level?}\antonio{ I dont think it replaces it, you could still have cases that gravitational significance, but not matter is needed.}

The dual nature of the metric is explicit in the construction of the renormalized
stress-energy tensor. The renormalization subtracts the Hadamard parametrix
$H(x,x')$ of Eq.~\eqref{eq:Twick}, whose four-dimensional form is
\begin{equation}
H(x,x')=\frac{1}{8\pi^{2}}\left[\frac{\Delta^{1/2}(x,x')}{\sigma(x,x')}
   +\sum_{n=0}^{\infty}v_n(x,x')\,\sigma^{n}(x,x')\,\ln\!\big(\sigma(x,x')/\ell^{2}\big)\right],
\label{eq.propagator.singular}
\end{equation}
with $\sigma(x,x')$ denoting Synge's world function (half the squared geodesic distance),
$\Delta^{1/2}(x,x')$ the square root of the Van Vleck--Morette determinant,
$\ell$ an arbitrary length scale, and $v_n$ the Hadamard coefficients. The
parametrix is constituted by two ingredients, in both of which the metric figures
in a purely spatio-temporal capacity. The first is the geometry of geodesic
distances: the world function $\sigma$ and the Van Vleck determinant, fixed by
the metric alone and independent of the field's mass or couplings. The second is
the field equation itself: the coefficients $v_0(x,x'),v_1(x,x'),\dots$ are fixed, order by order, by the Hadamard recursion relations that
follow from requiring $H$ to be a bisolution of $P\hat\phi=0$. In neither
ingredient does the metric do more than supply the background geometry -- the
geodesic distances along which the singularity propagates, and the wave operator
whose transport equations generate the $v_n$. At the level of the parametrix,
then, the metric is exhausted by its spatio-temporal role, and the $v_n$ are
local, covariant by-products of the geodesic distance and the wave operator.

The material significance appears only at the next step: renormalization delivers the local term $C_{\mu\nu}(x)$, and here the
\emph{same} datum acquires an energetic meaning: the coincidence limit
$v_1(x,x)$ is what fixes
the state-independent contribution to the conserved stress-energy -- it is (up to
a fixed factor, $\langle{:}\hat T^{\mu}{}_{\mu}{:}\rangle \propto
v_1(x,x)$) the trace anomaly \eqref{eq:anomaly} of the previous section. %The length scale $\ell$ left free in the logarithm is, moreover, exactly the scale whose rescaling $g_{\mu\nu}\to\lambda g_{\mu\nu}$ produced the $\ln\lambda$ of the Dimension argument: the very object that carries the energy also carries the anomalous scaling. 
The two significances are therefore carried by one and the same object, and neither
is prior to the other: the coefficient $v_1$ is at once a piece of spatio-temporal
structure -- built from geodesic distances and the field equation -- and a genuine
contribution to the stress-energy that gravitates. One cannot say what is subtracted
without reading $g_{\mu\nu}$ as spacetime, nor state the result of the subtraction
without reading $g_{\mu\nu}$ as a source of energy; the geometrical and the material
readings stand or fall together. It is this joint, non-hierarchical character -- a
single datum that is geometric in what it is built from and material in what it
contributes -- that underwrites the weak-egalitarian claim: in quantum field theory
on a curved background the metric carries geometrical, gravitational, and now material
significance on a par.
%A defender of the classical distinction might object that $C_{\mu\nu}[g]$ is merely a \rev{renormalization}{renormalization} counterterm and therefore need not be accorded any material significance. This objection overlooks both its physical role and its structural indispensability. Physically, $C_{\mu\nu}[g]$ contributes to the conserved stress-energy sourcing the semiclassical Einstein equation and thus back-reacts on the geometry; in the conformal case, this is reflected in the trace anomaly \eqref{eq:anomaly}. The ambiguity $\Theta_{\mu\nu}$ can shift some local curvature terms into the gravitational couplings, but cannot eliminate the anomalous contribution altogether. Structurally, $C_{\mu\nu}[g]$ is not an optional term appended to the theory: it is introduced precisely in order to obtain the required conservation law. Moreover, unlike the coupling to $J$, the coupling to the metric is universal, so a metric-dependent contribution of this kind enters the energetic balance of any quantum matter theory on curved spacetime. Quantum field theory thus has a twofold effect: it relaxes the assumptions that classically excluded a stress-energy contribution for the metric, as shown in §\ref{sec:quantum}, while at the same time tying that contribution irreducibly to the presence of quantum matter. The metric therefore satisfies the weak-egalitarian condition along a third axis: it carries geometrical, gravitational, and material significance together, the last emerging only once fields on the background are \rev{quantized}{quantised}.
\subsection{Unificationist view on gravitational and non-gravitational energy}
\label{sec:unificationist}

A second, historically grounded posture comes from Einstein's own reading of the
theory \citep{LEHMKUHL2014316}. Against the view that GR geometrizes gravity -- a
statement Einstein dismissed as ``saying nothing at all'' (\emph{nichtssagend}) --
Lehmkuhl argues that what Einstein prized was \textit{unification}: inertia and
gravity as two aspects of a single inertio-gravitational field, on the model of the
electric and magnetic fields in special relativity. The decomposition of the geodesic
equation into an ``inertial'' and a ``gravitational'' term is coordinate-dependent,
just as the split of $F_{\mu\nu}$ into $E$ and $B$ is frame-dependent. We abstract from this a criterion\footnote{This is our interpretation of the unificationist view, not the position held in \citep{LEHMKUHL2014316}.}.
\begin{interpretation}[Unificationist]
Two (or more) concepts are unified by a theory if:
\begin{enumerate}
\item[(i)] they can be separated by the presence of some given structure;
\item[(ii)] the theory provides no unique procedure for choosing that structure.
\end{enumerate}
\end{interpretation}

The electromagnetic field is the paradigm: relative to a timelike vector field
$u^\mu$, via $E_\mu=F_{\mu\nu}u^\nu$ and $B_\mu=-\tfrac12\varepsilon_{\mu\nu\alpha\beta}
u^\nu F^{\alpha\beta}$, the Faraday tensor always splits, but never invariantly,
because the theory privileges no $u^\mu$; the invariant is $F_{\mu\nu}$ and its
scalars. Einstein's inertia--gravity case has the same shape, with a coordinate
system as the free structure. It is this schema -- one object, decomposed only
relative to a structure the theory leaves free -- that we transpose to energy: we claim that gravitational and
non-gravitational energy are unified according to it.

The unification rests on two features of the semiclassical construction, one from
gravity and one from quantum theory. \emph{Universality}: every field couples to the
metric through its kinetic term, so the wave operator $P$, the Hadamard parametrix
$H$ it fixes, and the counterterm $C_{\mu\nu}(x)$ are all metric-built and present
for every sector. Unlike a background source $J$, whose counterterm is governed by a
coupling and vanishes as $\alpha\to0$, the metric admits no such switch:
$C_{\mu\nu}(x)$ remains on any curved background. \emph{Divergence}: the point-split
$\langle\hat\phi(x)\hat\phi(x')\rangle_\omega$ and the parametrix $H(x,x')$ are each
singular as $x'\to x$, and only their difference is finite. The field's stress-energy
and the geometric subtraction are thus not two finite objects summed but two
divergent objects whose only finite content is their combination; classically
$\phi^2$ is finite, no subtraction is needed, and the matter stress-energy stands
alone. It is quantization that fuses matter and geometry into a single finite object.

Granting the fused whole, can a vacuum recover the split? What subtracting a preferred
state delivers is not a gravitational/non-gravitational division but an
excitation/vacuum-polarization one,
\begin{equation}
 \underbrace{\big(\langle\wick{\hat T_{\mu\nu}}\rangle_\omega
   -\langle\wick{\hat T_{\mu\nu}}\rangle_{\mathrm{vac}}\big)}_{\text{matter (excitations)}}
 \;+\;
 \underbrace{\big(\langle\wick{\hat T_{\mu\nu}}\rangle_{\mathrm{vac}}
   \big)}_{\text{vacuum polarization}} .
\end{equation}
The first term is genuinely matter: in the difference the parametrix $H$, the
counterterm $C_{\mu\nu}$ and the curvature-induced vacuum energy all cancel, leaving
only how far the state departs from the vacuum. The second is \emph{not}
gravitational. The vacuum polarization is the response of the matter vacuum to the background, and
depends jointly on the curvature and on the matter field's own mass and coupling---and,
if further classical backgrounds are present, on those too. It is in general a mixture
rather than a clean gravitational energy; only for a massless, conformally coupled field
on a pure-metric background does it reduce to the purely geometric quantity of the
example below, though a non-vanishing curvature contribution always survives.

Nevertheless, a preferred vacuum is a strong demand. As $E$ and $B$ need a timelike $u^\mu$
and inertia and gravity a chart, this split needs additional structure selecting a preferred reference state. While in the former cases the required structure can always be chosen, here it must be supplied by symmetries of the spacetime geometry, such as a timelike Killing field, providing a preferred notion of positive frequency and a natural ground state, and is generically absent. The analogy is thus sharpened: the reference structure that would permit a split need not be available, and generically is not.

None of this makes the total ill-defined. As in electromagnetism the invariant is the
whole: $\langle\wick{\hat T_{\mu\nu}}\rangle$ is a well-defined, conserved, gravitating
tensor with a scheme- and state-independent invariant, e.g. the trace anomaly
\citep{Ferrero:2023unz}, playing the role of $E^2-B^2,\ E\cdot B$ and of the Riemann
curvature. What is arbitrary is only the division into sectors, and experiment does
not remove it: measurement fixes the total renormalized couplings, never the split of
the observed cosmological term into a bare gravitational and a matter vacuum-energy
contribution. 

The criterion fixes when two concepts are unified but not what unification does to
their status, and here two readings diverge. On \emph{weak unificationism}
the concepts remain individually meaningful, each well defined once the structure is
chosen---as $E$ and $B$ are, relative to an observer---so that only their
separation is convention-relative. On \emph{non-applicability unificationism} the
separate concepts do not survive at all: only the whole is a genuine quantity, and
the ``components'' are artefacts of a choice behind which there is no fact of the
matter. The two readings will give opposite verdicts on whether gravity is matter.
 
 On a weak
unificationist interpretation, both non-gravitational and gravitational contributions are meaningful. Let us see an example. One of the main components of our $\Lambda$CDM model is radiation. As a toy model it can be described by a conformally coupled massless scalar\footnote{See footnote~\ref{fn:conformal}.} in a spatially flat FLRW universe. The conformal symmetry allows us to choose a preferred vacuum state. Taking this as the reference state, the renormalized stress-energy tensor for a KMS state
\citep{Dappiaggi:2010gt} can be decomposed as 
\begin{equation}
  \underbrace{\langle{:}\hat T_{\mu\nu}{:}\rangle_{\mathrm{vac}}
 }_{\text{gravitational}}=-\frac{1}{2880\pi^{2}}\Big[\tfrac16\,{}^{(1)}\!H_{\mu\nu}-{}^{(3)}\!H_{\mu\nu}\Big],
\end{equation}
built from the two local, conserved curvature tensors\footnote{Here $^{(3)}\!H_{\mu\nu}$ is also a tensor constructed out of products of curvature but in contrast to $^{(1)}\!H_{\mu\nu}$ cannot be derived from a local action and therefore cannot be eliminated by an appropriate choice of $\Theta$.} and 
\begin{equation}
  \underbrace{\langle{:}\hat T_{\mu\nu}{:}\rangle_{\omega}
              -\langle{:}\hat T_{\mu\nu}{:}\rangle_{\mathrm{vac}}}_{\text{non-gravitational}}
  =(\rho_{\beta}+p_{\beta})\,u_{\mu}u_{\nu}+p_{\beta}\,g_{\mu\nu},
  \qquad
  \rho_{\beta}=\frac{\pi^{2}}{30\,(a\beta)^{4}},\quad p_{\beta}=\tfrac13\rho_{\beta},
\end{equation}
a traceless thermal-radiation fluid ($u^{\mu}u_{\mu}=-1$ the comoving four-velocity) that merely
redshifts as $a^{-4}$, with proper temperature $T(t)=1/\!\big(a(t)\beta\big)$. The non-gravitational component is the usual radiation but it comes with a vacuum polarization contribution, which, albeit negligible for observations, is nonetheless a gravitational contribution to the total energy budget of our Universe. The two concepts, non-gravitational and gravitational, are in this sense completely meaningful. As such, gravitational energy would be as meaningful as the energy associated with the radiation component of our Universe. It follows that, on this interpretation, gravity, i.e., the metric $g_{\mu\nu}$, would be considered matter at least as much as (quantum) photons (or any other particle) is considered matter.

\section{Conclusions and outlook}
\label{sec:conclusions}

% [keep Table~\ref{tab:summary} here]

We can now return to the question with which we began: what makes a quantum field a
matter field? Criterion~\ref{crit:2} is the minimal answer, and it assigns matter to
all the known relativistic fields---the fields of the Standard Model among them.
But applying it forces a further question the classical theory seemed to have settled:
is the metric $g_{\mu\nu}$ also a matter field?

Merely defining a stress-energy \emph{operator} for the quantum matter already
implicates the metric. The construction requires subtracting the Hadamard parametrix
$H(x,x')$ and adding the counterterm $C_{\mu\nu}(x)$, both built from the curvature:
the metric enters the very definition of the matter field's energy. This is the \emph{egalitarian} reading. On a curved
background the metric carries, alongside its spatio-temporal significance, a
\emph{material} one, and the two are borne by one and the same datum, neither prior to
the other. To that extent the identification of $g_{\mu\nu}$ with matter is supported.

Does the metric then have a stress-energy tensor of its own? Classically 
the theorem of \cite{Curiel:2018gfn} answered no. We argued that its underlying assumptions do not all carry over to the semiclassical theory. Its
dimension condition, which demanded a weight-zero object, is derived from the
scale-invariance of the Einstein equation; and in the semiclassical theory that
equation is \emph{not} scale-invariant, since the trace anomaly obstructs it. With the
weight-zero requirement gone, a local candidate $T^{\rm Grav}_{\mu\nu}$ can, after all,
be written.

Its status under the remaining requirement---separate conservation--- we read
deflationarily. The metric-built contribution does exchange energy with the
non-gravitational sector; it is, in that sense, a genuine participant in the balance of
stress-energy. But the total cannot in general be divided into a gravitational and a
non-gravitational part: universality makes the geometric term unavoidable, the
ultraviolet divergence welds it to the matter stress-energy, and a preferred vacuum---
the only structure that could separate them---exists only on backgrounds with enough
symmetry, and even there isolates not a gravitational energy but a mixed vacuum
polarization. This is the \emph{unificationist} reading: gravitational and
non-gravitational energy are unified, their separation fixed by no fact of the theory.

On weak unificationism the metric carries energy and is matter; on non-applicability unificationism the separate energies are meaningless, so gravity is not matter -- but neither are the quantum fields, whose energetic content is equally undefined without a vacuum, and matter contracts to the classical fields alone. Either way the classical divide between matter and spacetime does not survive quantization.

In this chapter we have shown one of several examples QFT offers for understanding gravity, matter, and possibly spacetime. To finish, we offer here at least three interesting effects within the field that would deserve similar attention:

\begin{enumerate}
  \item \emph{Particle production.} Classically the amount of
  matter in a region is an observer-independent fact and spacetime an inert stage;
  quantum field theory overturns both, and along the seam we have traced. The metric
  affects quantum matter in two distinct capacities: through its geometric-causal structure
  alone---a bifurcate Killing horizon can give rise to thermal behavior, as in the Unruh and
  Hawking effects \citep{Wald:1995yp}---and through its non-adiabatic dynamics, which creates pairs by
  parametric amplification \citep{Birrell:1982ix} exactly as a time-dependent external field does, the
  gravitational counterpart of the Schwinger effect realized in cosmological particle
  creation. The same phenomenon, the creation of matter, thus flows from the metric read
  once as spacetime structure and once as a matter-like external field.

    \item \emph{Radiating gravity.}
  Gravitational waves are the paradigm carriers of gravitational energy: classically a
  wave transports energy across null infinity through the Bondi flux, and it has been
  argued that this radiative energy is as real as its electromagnetic counterpart
  \cite{Gomes:2023xda}. Yet the idealized plane gravitational wave is invisible to the
  semiclassical stress-energy. Its curvature is type-N, so every quadratic scalar
  vanishes and the local geometric term is zero, $C_{\mu\nu}=0$; and, more than
  that, a plane wave does not polarize the vacuum \citep{Deser:1975ffb} and creates no
  particles \citep{Gibbons:1975jb}, so the full renormalized $\langle{:}\hat
  T_{\mu\nu}{:}\rangle_{\rm vac}$ vanishes as well. The same object through which the
  metric acquired a material significance thus registers \emph{nothing} for the
  configuration associated with gravitational energy in the classical
  theory. That a wave can carry energy while contributing zero to the quantum
  stress-energy---its local geometric \emph{and} its vacuum-polarization parts alike---
  is a genuine tension between the local semiclassical description and the quasi-local
  or non-local energy of radiation, and we must leave it as an open problem\footnote{The
  blindness is proper to the vacuum and to the plane-wave idealization: a generic
  radiative background does polarize the vacuum, and the excitations themselves still
  feel the wave, as the modified light propagation below shows}.

  \item \emph{Drummond-Hathrell effect.} The causal structure of a
  classical spacetime is standardly identified with the null cones of the metric.
  Quantum vacuum polarization breaks the identification: computing the one-loop QED
  effective action on a curved background, \cite{Drummond:1979pp} find that quantum vacuum-polarization effects introduce corrections controlled by the electron Compton wavelength, so that photons no longer propagate exactly on the null cones of $g_{\mu\nu}$ -- their propagation becomes polarization-dependent (a gravitational birefringence), is sensitive to tidal curvature, and may exhibit superluminal phase velocities. The effective causal structure seen by light is then not the metric's but a modified, matter-dependent one.
\end{enumerate}

\bibliographystyle{psalike}
\bibliography{citations} % Entries are in the citations.bib file

\end{document}